# Emergence of Nearly Flat Bands through an Embedded Kagome Lattice in an Epitaxial Two-dimensional Ge Layer on ZrB$_2$(0001)

A. Fleurence[1], C.-C. Lee[2*], R. Friedlein[1,**], Y. Fukaya[3], S. Yoshimoto[2], K. Mukai[2], H. Yamane[4,***], N. Kosugi[4,****], J. Yoshinobu[2], T. Ozaki[2], and Y. Yamada-Takamura[1]

*1. School of Materials Science, Japan Advanced Institute of Science and Technology (JAIST), Nomi, Ishikawa 923-1202, Japan*
*2. The Institute for Solid State Physics, The University of Tokyo, Kashiwa, Chiba 277-8581 Japan*
*3. Advanced Science Research Center, Japan Atomic Energy Agency, 2-4 Shirakata, Tokai, Naka, Ibaraki, 319-1195, Japan*
*4. Institute for Molecular Science, Okazaki, Aichi 444-8585, Japan*
*[*]Current affiliation: Department of Physics, Tamkang University, Tamsui, New Taipei 25137, Taiwan.*
*[**] Current affiliation: Meyer Burger (Germany) GmbH, 09337 Hohenstein-Ernstthal, Germany*
*[***] Current affiliation: RIKEN SPring-8 Center, Kouto, Sayo, Hyogo 679-5148, Japan*
*[****] Current affiliation: Institute of Materials Structure Science, KEK, Tsukuba 305-0801, Japan*

# Abstract

Ge atoms segregating on zirconium diboride thin films grown on Ge(111) were found to crystallize into a two-dimensional bitriangular structure which was recently predicted to be a flat band material. Angle-resolved photoemission experiments together with theoretical calculations verified the existence of a nearly flat band in spite of non-negligible in-plane long-range hopping and interactions with the substrate. This provides the first experimental evidence that a flat band can emerge from the electronic coupling between atoms and not from the geometry of the atomic structure.

Non-dispersive, or "flat bands", are peculiarly strongly-correlated electronic structures and can give rise to exotic topological states of matter [1-4] and electronic instabilities including ferromagnetism [5-9], Wigner crystallization [10] and superconductivity [11,12]. Flat bands can emerge as solution of the Schrödinger equation for Bloch tight-binding Hamiltonians of particular geometrically frustrated two-dimensional (2D) structures such as Lieb, checkerboard, dice or kagome lattices [13-15]. However, such 2D lattices remain rare in their free-standing forms, and are mostly observed within the crystal or at the surface of layered materials [9,16-18]. For instance, a nearly flat band and long-range ferromagnetic order have been reported for the layered, quasi-2D $Fe_3Sn_2$ crystal consisting of Fe kagome lattices [9]. Owing to novel functionalities they are expected to host, efforts were dedicated to design and fabricate 2D materials exhibiting electronic flat bands in their band structures [4,19-22]. As such, it has recently been demonstrated that Lieb, checkerboard, or kagome lattices can be embedded into not yet explored 2D structures [23,24]. Among them, a so-called "bitriangular" structure consisting of a hexagonal array of atoms overlaid with a ($\sqrt{3} \times \sqrt{3}$) array of atoms sitting on hollow sites of the former has been predicted to be capable of replicating the electronic band structure associated with kagome lattices [24]. For a particular relation between the difference of site energies and the hopping integral coefficients, one of the eigen solutions of the tight-binding Hamiltonian is a flat band [24].

In this Letter, we demonstrate that Ge atoms crystallize spontaneously into such a bitriangular lattice on Zr-terminated zirconium diboride ($ZrB_2$) thin films grown on the Ge(111) substrate surface. Furthermore, the electronic band structure as revealed by angle-resolved photoelectron spectroscopy (ARPES), in combination with density functional theory (DFT) calculations indicates that the condition for the emergence of the flat band is nearly fulfilled.

Epitaxial ZrB$_2$ thin films were grown by thermal decomposition of Zr(BH$_4$)$_4$ at the Ge(111) substrate surface kept at 650°C using a ultrahigh vacuum – chemical vapor epitaxy (UHV-CVE) system equipped with reflection high-energy electron diffraction (RHEED). Prior to growth, natural oxide was removed by heating the Ge substrate overnight at 650°C under UHV conditions. Epitaxial, single-crystalline ZrB$_2$ thin films grown on Ge(111) substrates adopt an epitaxial relationship with the substrate ZrB$_2$(0001)//Ge(111) and ZrB$_2$[11$\bar{2}$0]//Ge[1$\bar{1}$0], owing to a 4:5 lattice mismatch between the substrate and the thin film [25]. This is similar to the case of the Si(111) wafer as a substrate [26]. The difference is that, due to this epitaxial condition, the measured in-plane lattice constant of ZrB$_2$(0001), 3.157 Å, is slightly smaller than that for the bulk, while it was slightly larger in the case of Si(111) as a substrate [26]. Core-level spectra were recorded at the beamline BL-13 of KEK Photon Factory synchrotron radiation using a hemispherical electron energy analyzer (SCIENTA SES200). Angle-resolved photoemission spectra were measured at the beamline BL6U of the UVSOR-III Synchrotron at the Institute for Molecular Science. The overall energy and momentum resolutions were better than 20 meV and 0.03 Å$^{-1}$, respectively. Scanning tunneling microscopy (STM) was carried out in constant current mode using a JEOL UHV-STM and Pt-Ir tips. Total-reflection high-energy positron diffraction (TRHEPD) experiments were carried out at the Slow Positron Facility of the Institute of Materials Structure Science. The details were described elsewhere [27]. The beam energy of the incident positron was set at 10 keV. The rocking curves were measured by changing the glancing angle of the incident positron from 0.5 to 6.0° at a step of 0.1°. The intensity calculations were performed on the basis of the dynamical diffraction theory [28]. The structural and non-structural parameters, *i.e.*, the Debye temperature ($\Theta_D$), the adsorption potential due to the electronic excitations ($V_{el}$), and the mean inner potential ($V_0$) were optimized such as to minimize the difference between the experimental and the calculated curves. The goodness of fit was judged *via* the reliability (*R*) factor [29]. All experiments were carried out at room temperature. The first-principles calculations in the framework of DFT within the generalized gradient approximation (GGA) were performed using the OpenMX code [30], where the optimized pseudo-atomic basis functions were adopted. Two, two, and one optimized radial functions were allocated for the *s*, *p,* and *d* orbitals, respectively, for the Ge atoms with a cutoff radius of 7 Bohr, denoted as Ge7.0-*s2p2d*1. Zr7.0-*s3p2d*2 and B7.0-*s2p2d*1 were adopted for the Zr and B atoms, respectively. A cutoff energy of 110 Hartree was used for numerical integrations and for the solution of Poisson equation. A 6 × 6 k-point sampling was adopted for $ZrB_2(0001)$-($\sqrt{3} \times \sqrt{3}$) unit cell. A slab composed of 9 Zr and 8 B layers with Ge atoms locating on both sides were considered with the vacuum thickness of 20 Å. The structure was relaxed at theoretical lattice constant of bulk $ZrB_2(0001)$ until the atomic forces become less than 3 × 10$^{-4}$ Hartree/Bohr. For core-level binding-energy calculations [31], a slab with a $ZrB_2(0001)$-($3\sqrt{3} \times 3\sqrt{3}$) unit cell and composed of 5 Zr and 4 B layers, was adopted.

Figure 1(a) shows core-level spectra in the B1*s* and Zr3*d* region recorded before and after annealing at 770°C of a $ZrB_2$/Ge(111) sample under UHV. The spectrum recorded for the as-loaded sample (dark curve) shows the presence of Zr, B, and Ge oxides. In a manner reminiscent to $ZrB_2$ thin films grown on Si(111) [32], the native oxide layer can be removed by annealing under UHV conditions at temperatures in the 750-800°C range, as evidenced by the vanishing of the oxide-related peaks after annealing. Whereas the B1*s* peak shows a single component, the Zr3*d* signal can be decomposed into two components consisting of spin-split doublets, from which one can conclude that the ZrB$_2$ thin film is Zr-terminated (red spectrum). While the doublet with the Zr3$d_{5/2}$ peak at 178.89 eV can be assigned to bulk ZrB$_2$ [33], the one shifted by 160 meV to higher binding energies, corresponds to the Zr-terminated surface. The energy shift between bulk and surface components, significantly larger than the one observed for the (0001) surface of the bulk ZrB$_2$ material [33], is related to the (√3 × √3) reconstruction which is lacking for the bulk surface. A similar reconstruction was observed after the growth of ZrB$_2$ films on Ge(111) substrates when cooled down to temperatures below 450°C [25]. The formation of this reconstruction is apparently associated with the presence of Ge atoms segregating from the substrate after annealing as evidenced by the strong Ge3*d* peaks observed in the spectrum recorded under a surface-sensitive condition (Fig. 1(b)). The spontaneous segregation of Ge atoms from the substrate is reminiscent of ZrB$_2$ thin films grown on Si(111), on which segregated Si atoms crystallize as a honeycomb lattice ("silicene") resulting in the (2 × 2) reconstruction of the ZrB$_2$(0001) surface [34,35]. In contrast to the Si 2*p* spectrum of this epitaxial silicene layer [36], the spectrum of the Ge layer is notably simple as it appears as a single doublet of slightly asymmetric 3$d_{5/2}$ and 3$d_{3/2}$ peaks at 29.385 and 28.835 eV, respectively, having a full width at half maximum (FWHM) of 194 meV.

The ZrB$_2$(0001) thin film surface is composed of atomically flat terraces (Fig.1(c)) on which the Ge-rich reconstruction appears in STM images and LEED patterns as a (√3 × √3) array of protrusions (Fig.1(d)). The structure of the Ge layer was determined by fitting TRHEPD rocking curves, plotted by open circles in Fig. 2(a), with theoretical curves calculated for different structure models. Among them, a bitriangular structure shown in Figs. 2(b) and (c) for which the bottom atoms are sitting on hollow sites of the Zr-terminated ZrB$_2$(0001) thin film surface gives the best agreement with the experimental curves, as shown in Fig. 2(a). The best fitting was obtained for a distance of 2.3 Å between the topmost Zr layer and the bottom Ge layer and a distance of 1.85 Å between the protruding and bottom Ge layers. Note that the topmost Zr layer is buckled and shifted closer to the following B layer as compared to the distance in the bulk.

The structure obtained by TRHEPD was used as an input for DFT calculations with the purpose of finding an energetically stable structure. In the optimized structure shown in Figs. 3(a) and (b), bottom and top Ge atoms are respectively sitting at a distance of 2.3 Å and 4.1 Å above the topmost Zr layer, respectively. The calculated Ge3$d$ core-levels [30] are in agreement with the experimental spectrum as shown in Fig. 1(b) as the positions of the experimental peaks correspond well with the calculated binding energies of the bottom atoms. The difference of calculated binding energies between the two sites (0.15 eV) is smaller than the FWHM which makes the distinction between the two components difficult. Note that the spin-orbit splitting is slightly underestimated.

The agreement between the experimental results and the properties of the optimized structure is further demonstrated by the correspondences between the ARPES intensity plot (Fig. 3(b)) recorded along the $\Gamma$-$M$ direction of the Brillouin zone of the ZrB$_2$(0001) surface and the calculated band structure (Fig. 3(c)). The projection of the states on atomic orbitals indicates that the fringes centered at the $\Gamma$ point and the linearly dispersing bands crossing at the $M$ point originate mainly from ZrB$_2$ bulk states. The main contributions of Ge atoms can be found in the form of a band at a binding energy of 1.1eV that is dispersionless in the vicinity of the $\Gamma$ point, a shoulder in the binding energy range of 0.7 to 1.2 eV halfway in between the $\Gamma$ point and the $K$ point, and a band found between the Fermi level and a binding energy of 0.4 eV that is slightly dispersing around the $K$ point. The theoretical evaluation shows that most of the contributions to Ge-related bands are from the Ge $p_z$ orbitals.

To understand the origin of these Ge-related bands straightforwardly, we introduced a simplified tight-binding model described in the inset of Fig. 4(a). It is similar to the one discussed in Ref.[24]. In this model, a single orbital per atom with an isotropic in-plane symmetry (s, $p_z$,...) is considered. $t_1$ is the hopping integral between bottom atoms, $t_2$ is the hopping integral between top and bottom atoms, and $\varepsilon$, the difference of site energies between bottom and top atoms. As shown in Fig. 4(a), the band structure solution of the Schrödinger equation features a perfectly dispersionless unoccupied (pointed by a green arrow) if the condition (i) $t_1^2=\varepsilon t_2+3t_2^2$ is fulfilled [24].

The band structure of the free-standing Ge bitriangular lattice as isolated from the ZrB$_2$ substrate was calculated by DFT and plotted in Fig. 4(b), in which only the contributions of the Ge $p_z$ orbitals are shown. One can observe that the band structure stemming from the simple tight-binding model is essentially preserved. The nearly flat band (indicated by green arrow) in the 0.6-1.0 eV energy range above the Fermi level can be identified, suggesting that the condition (i) is almost fulfilled. However, the flat band turns into slightly dispersing bands with gaps due to hybridization with other Ge orbitals. On the other hand, the band indicated by the blue arrow looks very similar to the one obtained by tight-binding calculation.

Bringing the Ge bitriangular lattice in contact with the ZrB$_2$ surface causes charge transfer and hybridization between Ge and ZrB$_2$ states, and induces further modifications of the bands related to the 2D Ge layer and a shift to higher binding energies. A weak dispersing band crossing the Fermi level along the *K-M-K* direction (green arrow) seems to result from the hybridization of the flat band with the $ZrB_2$ surface band [37] and the shift in binding energy caused by electron transfer from the topmost $ZrB_2$ layer to the Ge bitriangular lattice. The band marked by a blue arrow also shifts to approximately 1 eV towards higher binding energies while keeping its shape. The electron transfer to the Ge lattice is consistent with the observed shift of surface-related $Zr3d$ peak component in core level spectrum.

Furthermore, as pointed out by the calculation of Mulliken charges, this charge transfer seems not to be homogeneous within the Ge layer. As shown in Table 1, once the Ge lattice is in contact with the Zr-terminated $ZrB_2$ surface, the charge transfer is three times higher for the bottom Ge atoms than for the top Ge atoms. The predominant contribution of the bottom Ge atoms to the flat band around the $\Gamma$ point [24] in combination with the charge transfer can explain the shift of that flat band (red arrow) towards higher binding energies of about 1 eV. This is observed in both calculations (Fig. 3(d) and Fig. 4(c)) and experiment (Fig. 3(c)).

In summary, we have experimentally demonstrated that an atomically thin Ge layer with a bitriangular structure forms spontaneously on $ZrB_2$ thin films grown on Ge(111) substrates. This layer manifests itself as a ($\sqrt{3} \times \sqrt{3}$) reconstruction of the $ZrB_2(0001)$ surface. The Ge bitriangular lattice adopts a 2D structure close to that required to give rise to a flat band [24]. This shows that it is possible to obtain flat bands from trivial structures. This flat band emerges from a particular balance between hopping integrals and site energies and not from the geometry of the atomic structure. It suggests that various electronic instabilities, as predicted to stem from a dispersionless band, can be easily tuned by modifying the structure, for instance, by means of adatoms or external strain.

The synchrotron radiation XPS experiments were performed at Photon Factory BL-13 under PF-PAC approval (No. 2009-S2-007). TRHEPD measurements were performed under the approval of the Photon Factory and Slow Positron Facility Program Advisory Committee of the Institute of Materials Structure Science (Proposal No. 2017G639 and No. 2019G684). ARPES measurements were supported by the Joint Studies Program (No. 615, 2012-2013) of the Institute for Molecular Science. Computational part of this work was carried out using the facilities in JAIST, supported by Nanotechnology Platform Program (Molecule and Material Synthesis) of the Ministry of Education, Culture, Sports, Science and Technology (MEXT), Japan. A.F. acknowledges support from the joint research program of the Institute for Solid State Physics, the University of Tokyo. Y. F. acknowledges financial supports from JSPS KAKENHI Grant Number JP18H01877 and a research grant from The Murata Science Foundation. Y. Y. T. acknowledges support from JSPS KAKENHI Grant Number JP20H00328.

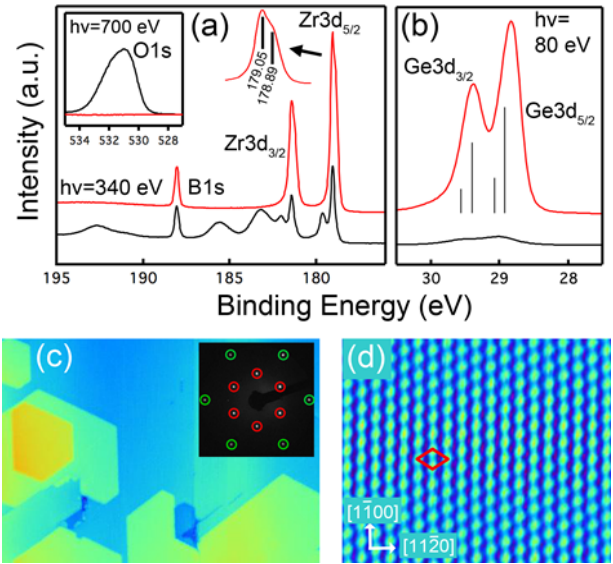

**Figure 1. Spontaneous segregation of Ge atoms on an oxide-free ZrB$_2$ thin film grown on a Ge(111) substrate.** (a) and (b): Photoelectron spectra recorded in the Zr3$d$-B1$s$ and in the Ge3$d$ core-level region, respectively, before (black) and after (red) annealing at 770°C for 2 hours. The inset in panel (a) shows spectra recorded in the O1$s$ region. Photon energies are indicated. The vertical bars at energies of 29.56 eV, 29.40 eV, 29.07 eV and 28.92 eV indicate calculated energies and the relative theoretical spectral weights of the Ge3$d$ peaks of the structure of Figs. 3(a) and (b). (c) and (d): STM images (150 × 100 nm$^2$, V=0.95 $V$, I=116 $pA$ and 5 × 4.3 nm$^2$, V=0.55 $V$, I= 56 $pA$) recorded after annealing. The (√3 × √3) unit cell is marked by a red rhombus. The inset shows LEED pattern recorded with electron energy of 70 eV. (1 × 1) and (√3 × √3) spots are indicated by green and yellow circles, respectively.

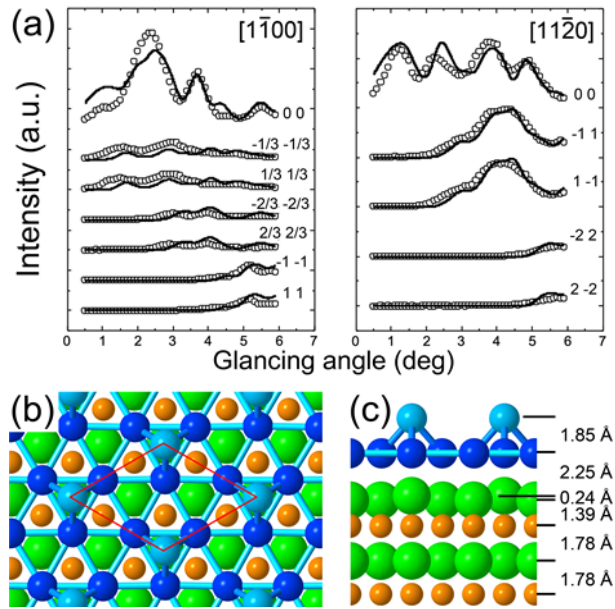

**Figure 2. The Ge bitriangular lattice on ZrB$_2$ thin films as determined from TRHEPD.** (a): TRHEPD rocking curves plotted with open circles and calculated curves as solid lines for various integer- and fractional-order diffraction spots (left: [1$\bar{1}$00] incidence, right: [11$\bar{2}$0] incidence]. (b) (c): Top and side views of the Ge bitriangular structure resulting from the fitting of the TREHPD rocking curves. Top and bottom Ge atoms are light blue- and blue-colored, Zr and B atoms are in green and orange, respectively. The red rhombus emphasizes the ZrB$_2$(0001)-($\sqrt{3} \times \sqrt{3}$) unit cell.

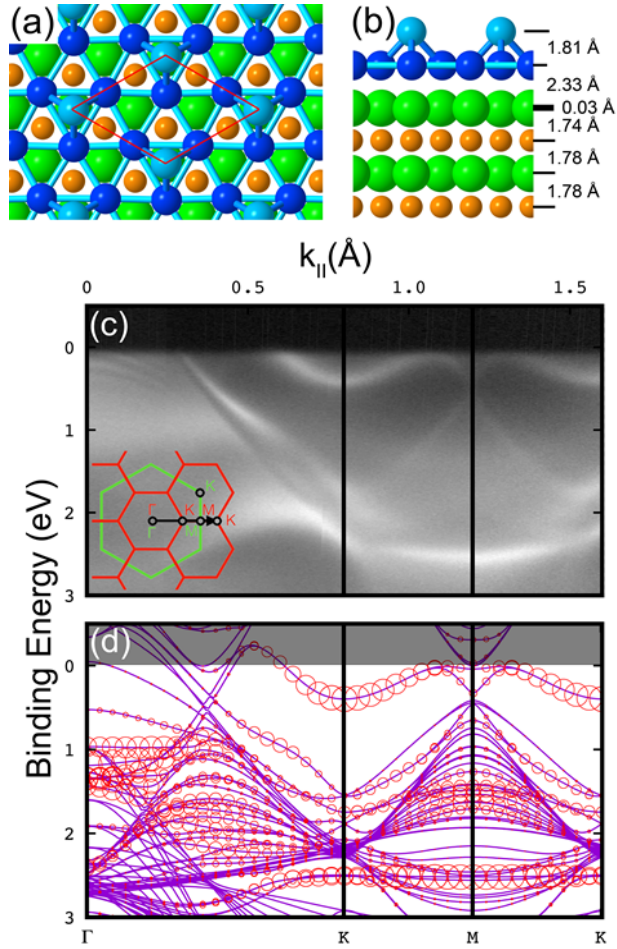

**Figure 3. Band structure of the Ge bitriangular lattice on ZrB$_2$.** (a) and (b): Top- and side-views of the bitriangular structure determined from DFT calculations. (c): ARPES intensity plot obtained from spectra recorded along the $\Gamma$-$M$ direction of the Brillouin zone of ZrB$_2$(0001)-(1×1) indicated by an arrow in the inset. Brillouin zones of the ZrB$_2$(0001)-(1×1) and of the bitriangular lattice are represented by green and red lines, respectively. (d) Calculated band structure of the Ge bitriangular lattice on ZrB$_2$(0001). The red fat bands show the contribution of the Ge p$_z$ orbitals.

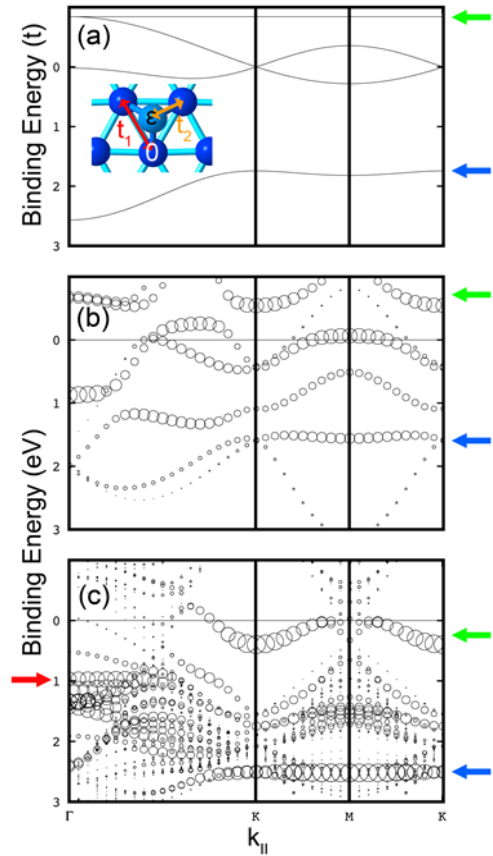

**Figure 4. Comparison of calculated band structures for the free-standing Ge bitriangular lattice and the one in contact with the ZrB$_2$ surface.** (a) Band structure originating from the tight-binding model shown in the inset for $t_1^2 = \varepsilon\, t_2 + 3 t_2^2$ with $t_1 = -2t$, $t_2 = -t$, and $\varepsilon = -t$. The green arrow at the side points towards the flat band while the blue arrow points towards the second band discussed in the text. (b) and (c): Contribution of the Ge p$_z$ orbitals to the band structures of free-standing Ge bitriangular lattice and of epitaxial Ge bitriangular lattice on ZrB$_2$(0001). Green and blue arrows in panels (b) and (c) indicate the bands that are derived from those indicated by the same arrows in panel (a). The red arrow points towards the flat band in the center of the Brillouin zone.

Table 1. Calculated Mulliken charges per Ge atom (expressed in number of electrons) for the top and bottom Ge atoms of the bitriangular Ge either with or without contact to the Zr-terminated $ZrB_2$ (0001) surface.

|  | **Free-standing Ge Layer** | **Ge layer on the $ZrB_2(0001)$ surface** | **Charge transfer** |
|---|---|---|---|
| **Top Ge atoms** | -0.074 | -0.113 | -0.039 |
| **Bottom Ge atoms** | 0.025 | -0.095 | -0.120 |